\documentclass[twocolumn,floatfix]{revtex4}
\usepackage{graphicx}
\usepackage{amsmath}
\usepackage{bm}
\begin{document}

\title{Accuracy of the diffusion equation with extrapolated-boundary condition for transmittance of light through a turbid medium}
\author{J. C. J. Paasschens}
\affiliation{Philips Research Laboratories, 5656 AA Eindhoven, The Netherlands}
\author{M. J. M. de Jong}
\affiliation{Philips Research Laboratories, 5656 AA Eindhoven, The Netherlands}
\author{C. W. J. Beenakker}
\affiliation{Instituut-Lorentz, Universiteit Leiden, P.O. Box 9506, 2300 RA Leiden, The Netherlands}
\begin{abstract}
The linear intensity profile of multiply scattered light in a slab geometry extrapolates to zero at a certain distance beyond the boundary. The diffusion equation with this ``extrapolated boundary condition'' has been used in the literature to obtain analytical formulas for the transmittance of light through the slab as a function of angle of incidence and refractive index. The accuracy of these formulas is determined by comparison with a numerical solution of the Boltzmann equation for radiative transfer.\medskip\\
{\em manuscript from 1995, published in J. C. J. Paasschens, Ph.D. Thesis (Leiden University, 1997)}
\end{abstract}
\maketitle

\section{Introduction}
\label{intro}

Multiple scattering of light in a turbid medium is well described by the
theory of radiative transfer \cite{EChan,EIsh,EHulst}. This theory is based on a
Boltzmann equation for the stationary intensity $I(\vec r,\hat s)$ of
monochromatic light at position $\vec r$ and with wave vector in the
direction $\hat s$. In the simple case of isotropic and non-absorbing
scatterers (with mean free path $l$), the Boltzmann equation takes the
form
\begin{equation}
  l\,\hat s\cdot \vec\nabla I(\vec r,
  \hat s) = -I(\vec r,\hat s) + \bar I(\vec r).
  \label{eq:Boltz1}
\end{equation}
Far from boundaries the angle-averaged intensity, given here for
$3$ di\-men\-sions, 
\begin{equation}
  \bar I(\vec r) \equiv \int \frac{d\hat s}{4\pi} I(\vec r,\hat s)
\end{equation}
satisfies the diffusion equation
\begin{equation}
  \nabla^2 \bar I(\vec r) = 0,
  \label{eq:diffeq}
\end{equation}
which is easier to solve than the Boltzmann equation. The diffusion
equation breaks down within a few mean free paths from the boundary,
and one needs to return to the Boltzmann equation in order to
determine $\bar I(\vec r)$ near the boundaries.

A great deal of work has been done on the choice of boundary conditions for
the diffusion equation which effectively incorporate the non-diffusive
boundary layer \cite{EChan,EIsh,EGiovanelli,EVries89,EFreund92,ELuck94,EAronson93}. 
These studies have led to the so called ``extrapolated-boundary
condition''
\begin{equation}
  \bar I(\vec r) = -\xi\,\hat n\cdot\vec\nabla \bar I(\vec r),
  \label{eq:extbound}
\end{equation} 
where $\vec r$ is a point on the boundary and $\hat n$ is a unit
vector perpendicular to the boundary and pointing outwards. 
Equation~(\ref{eq:extbound}) implies that a linear density profile extrapolates
to zero at a distance $\xi$ beyond the boundary. The extrapolation
length $\xi$  is of the order of the mean free path.

In this paper we consider transmission through a slab of finite
thickness $L$. We compute the transmittance $T$, the ratio between the
incident flux and the transmitted flux, by solving the
Boltzmann equation numerically.
Previous work on this problem used the diffusion equation with
the extrapolated-boundary condition \cite{EKaplan93,ELi93,EHaskell94,EFantini94} 
(or an alternative large-$L/l$ approximation \cite{ELuck94}) to derive
convenient analytical formulas for the dependence of $T$ on $L$ and
$l$, on the refractive index of the slab, and on the angle of
incidence. It is the purpose of the present study to determine the
accuracy of these formulas, by comparison with the results from the
Boltzmann equation.
We generalize and extend a previous study by De Jong
\cite{EJong94} in the context of electrical conduction through a
disordered metal, where the issue of refractive-index mismatch and
angular resolution has not been considered.

\section{Calculation of the transmittance from the Boltzmann equation}

\begin{figure}[tb]
\centerline{%
\includegraphics[width=0.7\linewidth,angle=-90]{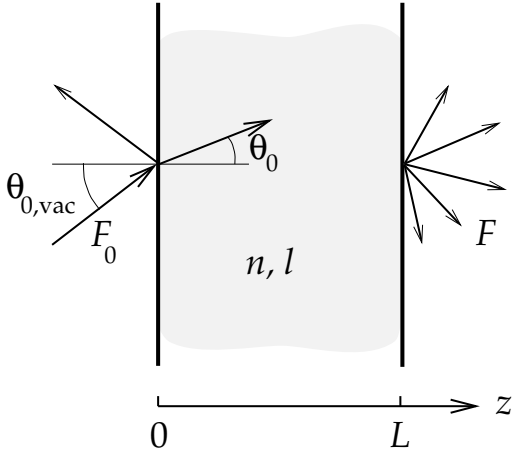}}
\medskip
\caption[Sketch of the slab geometry.]
{A sketch of the slab geometry.
The disordered medium (thickness $L$) has a refractive index $n$ (relative
to the outside) and a mean free path $l$. The
slab is illuminated by a plane wave incident at an angle $\theta_{0,\rm
vac}$, which is refracted to an angle $\theta_0$ inside the medium. The
transmittance $T$ is the ratio of the transmitted flux $F$ and the
incident flux $F_0$.}
\label{fig:geom}
\end{figure}

We consider a slab containing a disordered medium between the planes $z=0$
and $z=L$ (see Fig.~\ref{fig:geom}). The intensity $I(\vec r,\hat s)$
depends only on  the $z$-coordinate and on the angle $\theta$ between
$\hat s$ and the $z$-axis. We define $\mu\equiv\cos\theta$.
The Boltzmann equation~(\ref{eq:Boltz1}) takes the form
\begin{subequations}
  \label{eq:Boltz}
  \begin{eqnarray}
  l\mu\frac{\partial}{\partial z} I(z, \mu) &=& -I(z,\mu) + \bar I(z),
   \qquad\mbox{$0<z<L$,}\\
  \bar I(z) &=& \case12\int_{-1}^1 d\mu\, I(z,\mu).
  \end{eqnarray}
\end{subequations}%
We supplement Eq.~(\ref{eq:Boltz}) with  boundary conditions at $z=0$ and
$z=L$ that describe reflection due to a refractive index
mismatch, with reflection probability $R(\mu)$ :
\begin{subequations}
  \label{eq:BCboltz}
\begin{eqnarray}
  I(0,\mu)  =& R(\mu) I(0,-\mu) + I_0(\mu) ,\qquad&\mbox{$\mu>0$},\\
  I(L,-\mu) =& R(\mu) I(L,\mu) .
    \null\egroup$\hfill$\bgroup\qquad
     &\mbox{$\mu>0$},
  \end{eqnarray}
\end{subequations}%
The boundary condition at $z=0$ also contains the intensity due to a
planar source with angular distribution $I_0(\mu)$ inside the medium.
Note that the angular distribution is different from that outside the
medium, due to refraction at the boundary and due to the fact that
part of the light is reflected before even entering the medium.

The Boltzmann equation~(\ref{eq:Boltz}) with boundary
conditions~(\ref{eq:BCboltz}) implies for $\bar I(z)$ an integral equation
of the Schwarzschild-Milne type \cite{ELuck94,EFreund92,EChan,EJong94}
\begin{eqnarray}
  \bar I(z) =&  M_0(z)+\displaystyle\int_0^L d z'\,\bar I(z')\Bigl[
     M_1(z-z') +\nonumber\\
    &M_2(z+z') + M_2\bigl[(L{-}z) + (L{-}z')\bigr]+\nonumber\\
    &M_3(z-z') + M_3\bigl[(L{-}z) - (L{-}z')\bigr]\Bigr].
  \label{eq:MilneSlab}
\end{eqnarray}
We have defined the kernels
\begin{subequations}
\begin{eqnarray}
  M_1(z) &=& \int_0^1\frac{d\mu}{2l\mu}\,e^{-|z|/l\mu}, \\
  M_2(z) &=& \int_0^1\frac{d\mu}{2l\mu}
    N(\mu)
     R(\mu)\,e^{-z/l\mu},\\
  M_3(z) &=& \int_0^1\frac{d\mu}{2l\mu}
     N(\mu)
       R^2(\mu)e^{-(2L+z)/l\mu}.
\end{eqnarray}
\end{subequations}%
The factor $N$ is given by
\begin{equation}
  N(\mu) = \Bigl(1-R^2(\mu)e^{-2L/l\mu}\Bigr)^{-1}.
\end{equation}
The kernels $M_1$, $M_2$, and $M_3$ describe propagation from $z'$ to
$z$ with zero, an odd number, and an even number of reflections,
respectively.
The source term $M_0$ is given by
\begin{equation}
  M_0(z) =  \case12\int_0^1d\mu\,
  N(\mu)
   I_0(\mu)\;\Bigl[ e^{-z/l\mu} + R(\mu)e^{-(2L-z)/l\mu}\Bigr].
\end{equation}
Once $\bar I(z)$ is known, the intensities $I(z,\mu)$ and $I(z,-\mu)$
with $\mu>0$  follow from
\begin{subequations}
\begin{eqnarray}
  I(z,\mu) &=&
    I_0(\mu)N(\mu)e^{-z/l\mu}  +
    \int_0^z\frac{d z'}{l\mu} e^{(z'-z)/l\mu}\bar I(z')\nonumber\\
    &&\mbox{} +
    N(\mu) R(\mu) e^{-z/l\mu}
\nonumber\\&&\mbox{}\times
  \int_0^L\frac{d z'}{l\mu}
      \Bigl(e^{-z'/l\mu} + R(\mu)e^{-(2L-z')/l\mu}\Bigr) \bar I(z') ,\nonumber\\
&&         \\
  I(z,-\mu)&=&
    I_0(\mu) N(\mu) R(\mu) e^{-(2L-z)/l\mu}\nonumber\\
    &&\mbox{}+
    \int_z^L\frac{d z'}{l\mu}e^{(z-z')/l\mu}\bar I(z')
\nonumber\\&&\mbox{}+
    N(\mu) R(\mu) e^{-(2L-z)/l\mu}
\nonumber\\&&\mbox{}\times
    \int_0^L\frac{d z'}{l\mu}
      \Bigl(e^{z'/l\mu} + R(\mu)e^{-z'/l\mu}\Bigr) \bar I(z').
\end{eqnarray}
\end{subequations}
This is a solution of the Boltzmann equation~(\ref{eq:Boltz}) 
with boundary conditions~(\ref{eq:BCboltz}), as  can be
checked by substitution. Integration over all $\mu$ then yields the
Schwarzschild-Milne equation~(\ref{eq:MilneSlab}).
We solve the integral equation~(\ref{eq:MilneSlab}) numerically, by
discretizing the interval $(0,L)$, so that it reduces
to a matrix equation \cite{EJong94}.

The quantity of interest is the transmittance $T$, defined as the ratio
of the flux $F$ that is transmitted through the slab and the flux $F_0$
incident from the source,
\begin{equation}
  T = F/F_0.
\end{equation}
The transmitted flux is given by 
\begin{equation}
  F = 2\pi \int_{-1}^1d\mu\,\mu\,I(z,\mu),
\end{equation}
where $c$ is the speed of light in the medium. The flux is independent
of~$z$, because there is no absorption. The total
incident flux (including the flux which is reflected at the slab
boundary before entering the medium) is given by
\begin{equation}
  F_0=2\pi\int_{\mu_c}^1d\mu\,\frac{\mu I_0(\mu)}{1-R(\mu)}.
\end{equation}
We assume that the medium in the slab has a refractive index $n>1$
(relative to the refractive index outside the slab). The lower bound
$\mu_c$ in the integral, defined by  $\mu_c\equiv(1-1/n^2)^{1/2}$, is the
cosine of the angle at which total internal reflection occurs
[$R(\mu)\equiv 1$ for $\mu<\mu_c$].
For $0<\mu<\mu_c$ the reflection probability is given by the Fresnel formula
for unpolarized light, 
\begin{subequations}
\begin{eqnarray}
    R(\mu) &=& \case12\left|
       \frac{\mu_{\rm vac} - n\mu}{\mu_{\rm vac} + n\mu}
  \right|^2 + \case12\left|
     \frac{n \mu_{\rm vac} - \mu}{n \mu_{\rm vac} + \mu}
  \right|^2,\\
  \mu_{\rm vac} &\equiv& [1-n^2(1-\mu^2)]^{1/2}.
  \label{eq:Snell}
\end{eqnarray}
\end{subequations}%
The relation between $\mu_{\rm vac}=\cos\theta_{\rm vac}$ and
$\mu=\cos\theta$ is Snell's law, such that the angle of 
incidence $\theta_{\rm vac}$ outside the medium (in ``vacuum'') is 
refracted to an angle~$\theta$ in the medium (see Fig.~\ref{fig:geom}).  

\begin{figure}
\centerline{%
\includegraphics[width=0.9\linewidth]{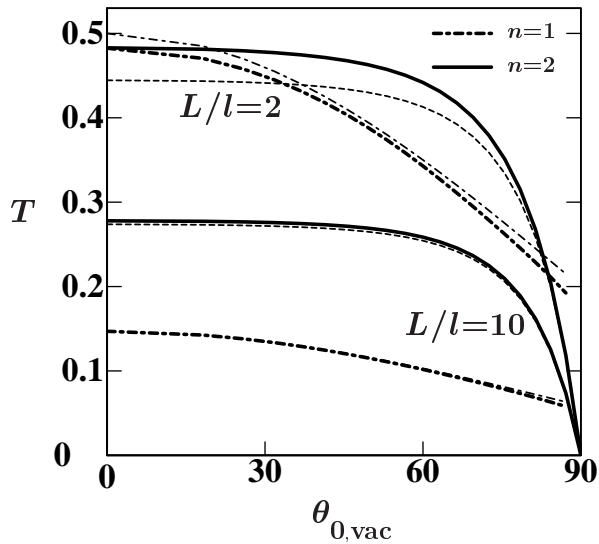}}
\medskip
\caption[Transmittance through a slab as a function of the incident
angle.]{
Transmittance as a function of the incident angle
$\theta_{0,\rm vac}$ for two different values of $L$ and $n$. The thick
curves are computed from the Boltzmann equation, the thin curves are the
diffusion approximation~(\protect\ref{eq:Tdiff}).}
\label{fig:Tmu0}
\end{figure}

We have calculated the transmittance for the case of  plane-wave
illumination, $I_0(\mu) = I_0\delta(\mu-\mu_0)$.  The results are shown in
Fig.~\ref{fig:Tmu0}, where $T$ is plotted  as a function of the 
angle of incidence $\theta_{0,{\rm vac}}$ outside the medium
[$\mu_{0,{\rm vac}}=\cos\theta_{0,{\rm vac}}$ is related to $\mu_0$ by
Eq.~(\ref{eq:Snell})]. We show results for two different ratios $L/l$ and
two values of $n$ (thick curves). 
 For $n=1$ the transmittance is non-zero for all incident angles,
but numerical difficulties prevent
us from going beyond $\theta_{0,{\rm vac}}\simeq 87^{\circ}$. The thin
curves in Fig.~\ref{fig:Tmu0} are the results of the diffusion approximation,
which we discuss in the following section.

\section{Comparison with the diffusion approximation}
The diffusion approximation for the transmittance has been studied by
several authors \cite{EKaplan93,ELi93,EHaskell94,EFantini94}. Here we briefly describe this approach, and
then compare the result with the numerical solution of the
Boltzmann equation.

In a slab geometry the diffusion equation with extrapolated-boundary
condition takes the form [cf.\ Eqs.~(\ref{eq:diffeq}) and~(\ref{eq:extbound})]
\begin{equation}
  \frac{d^2}{d z^2} \bar I(z) = 0,\qquad0<z<L,
  \label{eq:diffeq2}
\end{equation}
with boundary conditions
\begin{equation}
  \bar I(0) = \xi\bar I'(0),\qquad \bar I(L) = -\xi \bar I'(L).
  \label{eq:diffeq2BC}
\end{equation}
We assume plane-wave illumination of the boundary $z=0$, at an angle
$\theta_{0,{\rm vac}}$ with the positive $z$-axis. A fraction
$1-R(\mu_0)$ of the incident flux~$F_0$ enters the medium, and is first
scattered on average at $z=\mu_0 l$. [We recall that $\mu_0
=\cos\theta_0$, where $\theta_0$ corresponds to $\theta_{0,{\rm vac}}$
after refraction, cf.\ Eq.~(\ref{eq:Snell}).] This plane-wave
illumination is incorporated into the diffusion
equation~(\ref{eq:diffeq2}) as a source term,
\begin{equation}
  \frac{d^2}{d z^2} \bar I(z) + \frac 3{4\pi l}[1-R(\mu_0)] F_0
      \delta(z-\mu_0 l) =0.
  \label{eq:diffeq3}
\end{equation}
The solution of Eq.~(\ref{eq:diffeq3}) with boundary
condition~(\ref{eq:diffeq2BC}) is
\begin{equation}
  \bar I(z) =
  \begin{cases}
    \displaystyle\frac{3}{4\pi l}
                 \frac{(\xi+z)(L+\xi-\mu_0l)}{L+2\xi} &\cr
                 \quad\mbox{}\times[1-R(\mu_0)] F_0,&
              \mbox{if $0<z<\mu_0l$,}\cr
    \displaystyle \frac{3}{4\pi l} 
               \frac{(\xi+\mu_0l)(L+\xi-z)}{L+2\xi} &\cr
              \quad\mbox{}\times[1-R(\mu_0)] F_0,&
              \mbox{if $\mu_0l<z<L$.}
  \end{cases}
\end{equation}
The transmitted flux $F=-\frac{4}{3}\pi\bar I'(L)$ divided by the 
incident flux $F_0$
leads to the transmittance in the diffusion approximation,
\begin{equation}
  T_{\rm diff} = [1-R(\mu_0)] \frac{\xi + \mu_0 l}{L+2\xi}.
  \label{eq:Tdiff}
\end{equation}
This simple analytical formula combines results in the literature by
Kaplan et~al. \cite{EKaplan93} (who considered normal incidence) and by
Nieuwenhuizen and Luck \cite{ELuck94} (who considered the Schwarzschild-Milne
equation in the diffusive limit $L\gg l$).

\begin{figure}
\centerline{%
\includegraphics[width=0.9\linewidth]{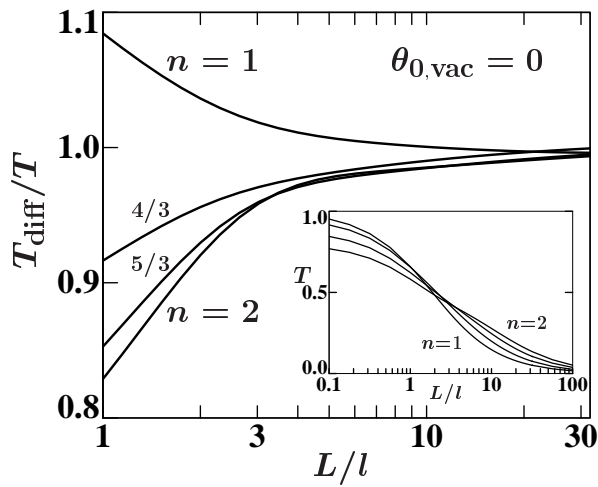}}
\medskip
\caption[Ratio of the transmittances from diffusion and the exact result.]{
Ratio of the transmittance $T_{\rm diff}$ according to the
diffusion approximation and the exact result $T$ according to the
Boltzmann
equation, for normal incidence
$\theta_{0,\rm vac}=\theta_0=0$. The inset shows $T$ as a function of
$L/l$, for the same values of $n$ as the main plot.}
\label{fig:Tmu0=1}
\end{figure}

We still need to specify the value of the extrapolation length $\xi$. We
will use an expression due to Zhu, Pine, and Weitz \cite{EZhu91},
\begin{equation}
  \xi = \case23 l\frac{1+C_2}{1-C_1},
  \label{eq:xiZhu}
\end{equation}
where the coefficients $C_1$ and $C_2$ are the first two moments of
$R(\mu)$,
\begin{subequations}
\begin{eqnarray}
  C_1 &=& 2\int_0^1d\mu\,\mu\,R(\mu),\\
  C_2 &=& 3\int_0^1d\mu\,\mu^2\,R(\mu),
\end{eqnarray}
\end{subequations}%
normalized such that $C_1=C_2=R$ for an angle-independent reflection
probability $R(\mu)=R$. Comparison of Eq.~(\ref{eq:xiZhu}) with a numerical
solution of the Boltzmann equation in a semi-infinite medium by
Aronson \cite{EAronson93} shows that it accurately describes the length
over which the linear density profile extrapolates to zero. The
difference is largest for $n=1$, when Eq.~(\ref{eq:xiZhu}) gives
$\xi=\case23l$ while the Boltzmann equation gives an extrapolation
length of $0.7104\,l$ which is somewhat larger \cite{EChan,ELuck94}.
The transmittance~$T_{\rm diff}$ is compared in Figs.~\ref{fig:Tmu0}
and~\ref{fig:Tmu0=1} with the exact $T$ from the Boltzmann equation.

Once the transmittance $T$ for plane-wave
illumination as a function of $\mu_{0,\rm vac}=\cos\theta_{0,\rm vac}$
is known, one can compute the transmittance $T_{\rm tot}$
for diffusive illumination by integrating over the angles of incidence,
\begin{equation}
  T_{\rm tot} = 2\int_0^1 d\mu_{0,\rm vac}\,\,
   \mu_{0,\rm vac}\, T(\mu_{0,\rm vac}).
\end{equation}
The diffusion approximation~(\ref{eq:Tdiff}) and~(\ref{eq:xiZhu}) yields the
analytical formula
\begin{equation}
  T_{\rm diff,tot} = n^2\left( \frac{3L}{4l} +
                            \frac{1+C_2}{1-C_1}\right)^{-1}.
  \label{eq:Tdifftot}
\end{equation}
In the absence of a refractive-index mismatch ($n=1$, $C_1=C_2=0$)
this formula has been found \cite{EJong94} to agree with the Boltzmann
equation within~3\% for all $L/l$.
For $n>1$ the relative error in Eq.~(\ref{eq:Tdifftot}) is comparable to
that shown in Fig.~\ref{fig:Tmu0=1} for the transmittance at normal
incidence.

In conclusion, we have computed the transmittance of a turbid medium of
mean free path $l$ and length $L$  from the Boltzmann equation as a
function of the angle of incidence. We compared the results from the
diffusion equation to this exact solution. The difference between the two
transmittances stays below 6\% for $L>3l$
and $1<n<2$. The diffusion approximation overestimates the transmittance
for $n=1$ and
underestimates it in the presence of a significant refractive index
mismatch. The relative error is largest for
large refractive index mismatch.

\acknowledgements
Discussions with G. W. 't Hooft are gratefully acknowledged. This work was supported by the Dutch Science Foundation NWO/FOM.

\end{document}